\documentclass[11pt,twoside]{article}
\usepackage{./asp2014}

\aspSuppressVolSlug
\resetcounters

\begin{document}

\title{Galactic Archaeology and Minimum Spanning Trees}
\author{Ben A. MacFarlane,$^1$ Brad K. Gibson,$^2$ and Chris M. L. Flynn$^3$
\affil{$^1$Jeremiah Horrocks Institute, University of Central Lancashire, Preston, UK}
\vspace{-2.0mm}
\affil{$^2$E.~A. Milne Centre for Astrophysics, University of Hull, Hull, UK}
\vspace{-2.0mm}
\affil{$^3$Centre for Astrophysics \& Supercomputing, Swinburne University, Australia}}

\paperauthor{Ben A. MacFarlane}{bmacfarlane@uclan.ac.uk}{ORCID_Or_Blank}{University of Central Lancashire}{Jeremiah Horrocks Institute}{Preston}{Lancashire}{PR1 2HE}{United Kingdom}
\paperauthor{Brad K. Gibson}{brad.gibson@hull.ac.uk}{ORCID_Or_Blank}{University of Hull}{E. A. Milne Centre for Astrophysics}{Hull}{East Riding of Yorkshire}{HU6 7RX}{United Kingdom}
\paperauthor{Chris M. L. Flynn}{cmlflynn@gmail.com}{ORCID_Or_Blank}{Swinburne University of Technology}{Centre for Astrophysics and Supercomputing}{Melbourne}{Victoria}{PO Box 218}{Australia}

\begin{abstract}
Chemical tagging of stellar debris from disrupted open clusters and 
associations underpins the science cases for next-generation 
multi-object spectroscopic surveys. As part of the Galactic 
Archaeology project TraCD (Tracking Cluster Debris), a preliminary 
attempt at reconstructing the birth clouds of now 
phase-mixed thin disk debris is undertaken using a parametric minimum 
spanning tree (MST) approach. Empirically-motivated chemical abundance 
pattern uncertainties (for a 10-dimensional chemistry-space) are 
applied to NBODY6-realised stellar associations dissolved into a 
background sea of field stars, all evolving in a Milky Way 
potential. We demonstrate that significant population reconstruction 
degeneracies appear when the abundance uncertainties approach 
$\sim$0.1~dex and the parameterised MST approach is employed; more 
sophisticated methodologies will be required to ameliorate these 
degeneracies.
\end{abstract}

\vspace{-4mm}

\section{Introduction}

\vspace{-4mm}
The underlying premise of Galactic Archaeology is that 
surveys provide a fossil 
record of the evolution of the Milky Way.  Mining this 
record entails the search for
sub-clustering in multi-dimensional (spatial, kinematic, chemical) 
datasets.  For systems with long dynamical times,
relatively few dimensions are needed to identify clustering; e.g.,
energy--angular momentum phase space alone can identify 
the building blocks of the stellar halo \citep{bk_03}.
Unfortunately, the dominant baryonic
component of the Galaxy - the thin disk - does not fall into this
somewhat `straightforward' regime.

Our thin disk is thought to have been built by many generations of 
now-disrupted stellar associations, the debris from which having
been subsequently scattered/migrated by a convolution of 
processes, including systematic
spiral arm- and bar-driven `churning', and random 
diffusion-like kinematic heating from giant molecular clouds.
Before an association has fully disrupted, 
identifying stellar siblings - i.e., the parent birth 
cloud/association - is relatively straightforward.  Spatial, kinematic, 
and/or phase space coherency can be maintained for a few $\sim$100~Myrs 
(depending upon cluster mass, concentration, and galactocentric 
radius/orbit).  Unfortunately, on the $\sim$10~Gyr timescale of the thin 
disk, the combined effects of diffusive scattering and radial migration 
quickly wash out this coherency, making sub-clustering analysis in 
low-order spatial and kinematic dimensions a fruitless endeavour (at 
least for reconstructing the birth locations of the sea of Galactic 
field stars and searching for our own Sun's siblings).

To combat this dimensionality `problem', \citet{fbh_02} proposed the use 
of 10-20 dimensions of chemistry-space (or `C-space').
Dubbed `chemical tagging', the principle hinges on the 
presumption that if the gas clouds from which the now-dissolved stellar 
associations formed were chemically homogeneous,
even after dissolution and full configuration- and phase-space 
mixing of the debris, the parent clouds' chemical `fingerprints' 
would remain invariant and identifiable.  This
presumption has been shown to hold 
empirically, with chemical homogeneity confirmed on an 
element-by-element basis (at the $\sim$0.1~dex level, spanning a range 
of nucleosynthetic processes) for $>$20 associations 
\citep{ds_07,ds_09}.

With homogeneity confirmed, the first `blind' chemical tagging 
experiments were conducted \citep{m_13,m_14}.  High-resolution 
spectroscopic data for field stars were analysed, to 
attempt a probabilistic approach to identifying cluster populations 
lurking in the field. Though the study presented a means to analyse the 
datasets from next-generation 
surveys, no means for explicitly determining cluster/association 
recovery percentages was presented.  Our goal within TraCD (Tracking 
Cluster Debris: \citealt{ml_15}) is to build on this pioneering work 
and characterise parametric and non-parametric approaches to 
multi-dimensional group finding within C-space, with the goal 
being the development of tools which can inform upcoming 
surveys such as GALAH, WEAVE, and 4MOST.

\vspace{-4mm}

\section{Method}

\vspace{-4mm}
As detailed by \citet{ml_15}, our
framework is a static 
3-component (logarithmic halo, Plummer sphere bulge, Miyamoto-Nagai
disk) potential; $\sim$10$^5$ disk stars, equally spaced
in ages up to 10~Gyrs old, are evolved with an N-body integrator
with treatments of both random molecular cloud
scattering and systematic spiral arm churning \citep{sb_02}
applied at each timestep. This background sea of stars
possess kinematics consistent with those of the
Milky Way.  We employ four
NBODY6 realisations of 250~M$_\odot$ stellar associations,
each evolved in the same 3-component
potential as the background stars; these
are injected into the potential at various galactocentric
radii $r_0$ ranging from 4 to 10~kpc; e.g., the dissolution time for a 
cluster injected at the solar circle (8~kpc) is
$\sim$0.5~Gyrs.  As stars escape the dissolving association, their
trajectories are tracked with the same integrator advancing the
positions of the background stars, and the same 
diffusion and churning treatments applied.

We tag our background field stars with empirical radial abundance
patterns drawn from \cite{ll_11}, using 10 dimensions
of C-space (Al ,Mg, Si, Ca, Ti, Cr, Co, Ni, Y, Nd). Using (of order) this
number of elements ($N_C$$\sim$10) provides the leverage to span
the breadth of nucleosynthesis sites,
while minimising the 
search through parameter space.  Because of the imposed (i) radial
abundance gradients, and (ii) constant
star formation history for our disk, the applied diffusion and churning
means that we consequently also impose a temporal evolution
pattern to the abundance gradient \citep{gp_13}.  The gradient's dependence
on time $t_0$ and radial position $r_0$ is used to tag the mean abundance
of the associations injected into the background
stars at any given time and location, with the user also imposing 
an element-by-element scatter $\sigma_C$ to each pre-disrupted system.
An important difference for the latter is that the
chemical `fingerprint' pattern imposed on the pre-disrupted system
is not `random', but instead unique and homogeneous, as per
\citet{ds_07,ds_09}.

Through use of a minimum spanning tree algorithm (MST - \cite{a_09}), 
we attempt to identify the debris of disrupted satellites amongst
the background stellar disk.  From the 10-dimensional C-space, the
MST determines a level of similarity $\delta_C$ 
between all single stellar components:
\begin{displaymath}
\delta_C = \sum_{C}^{N_C} \frac {| A_{C}^{i} - A_{C}^{j} |}{N_C},
\end{displaymath}
%\noindent
where $i$ and $j$ are respective stars and $A_C$ the 
abundance for element $C$ \citep{m_13}. In
order to deconvolve the matrix of $\delta_C$ values into likely parent
stellar clusters, MST
begins building a similarity tree. To
do so, stellar components of greatest $\delta_C$ value are 
joined as nodes, with subsequent iterations joining less similar stellar 
components until all stars are placed in the similarity tree. Having 
created the tree, a parametric exit condition is defined in 
which the similarity tree is pruned to a value of $\delta_C$ where the 
number of clusters, $k$, is present. Fig~\ref{Fig1} illustrates the 
building of a similarity tree for a toy distribution, in which it
is built (right panel) and subsequently pruned. Due to 
the nature of the similarity matrix, the parameterised pruning method 
may dissociate tree constituents. Such dissociation events thus have no 
cluster association.

\articlefiguretwo{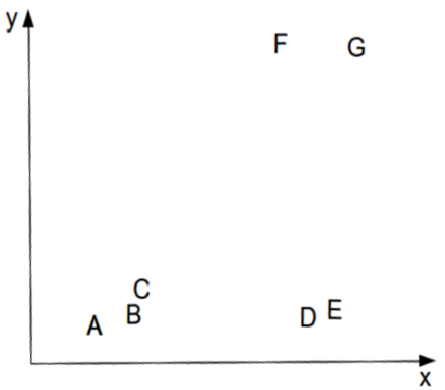}{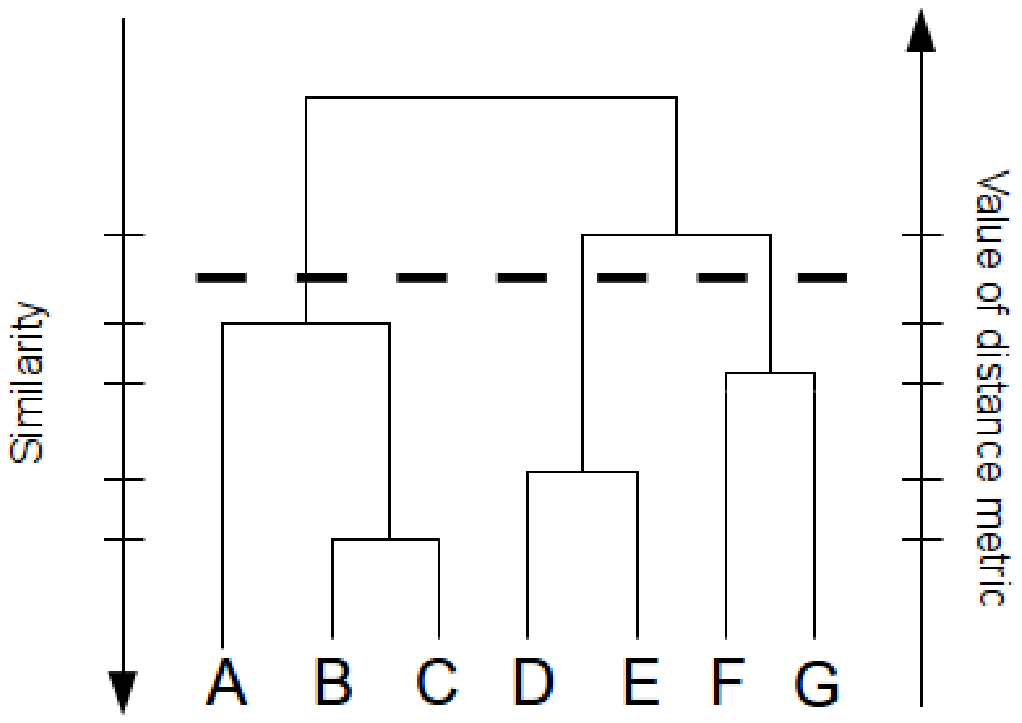}{Fig1}{Toy 
model illustrating three clustered populations in $x-y$ space (left).
MST builds a similarity tree, iteratively joining nodes 
of greatest $\delta_C$ until all components are included in the 
tree (right). A parametric cluster condition is then 
applied (dashed line) to define the cluster 
constituents.}

\vspace{-4mm}

\section{Results}

\vspace{-4mm}
As a proof-of-concept, we evolved four systems with
$r_0$ (kpc) = [4,6,8,10] for 5~Gyrs. To mimic observations,
we filter stars at the end of the simulation to only include
those within 3~kpc of our imposed solar neighbourhood (centred on 
$(x,y,z)$=($-$8.5,0,0)~kpc.
To determine the success of the MST cluster 
`reconstruction', four chemical abundance uncertainties
$\sigma_C$ (dex) were explored: [0.01,0.05,0.10,0.15].
Having tagged each star particle (see Fig~\ref{Fig2}), the MST 
was deployed, using a parametric exit condition of $k=4$. 
From comparisons of the `real' cluster constituents vs the MST 
associations, the successful cluster recovery rates and 
dissociation percentages were derived (see Tbl~1).

\articlefiguretwo{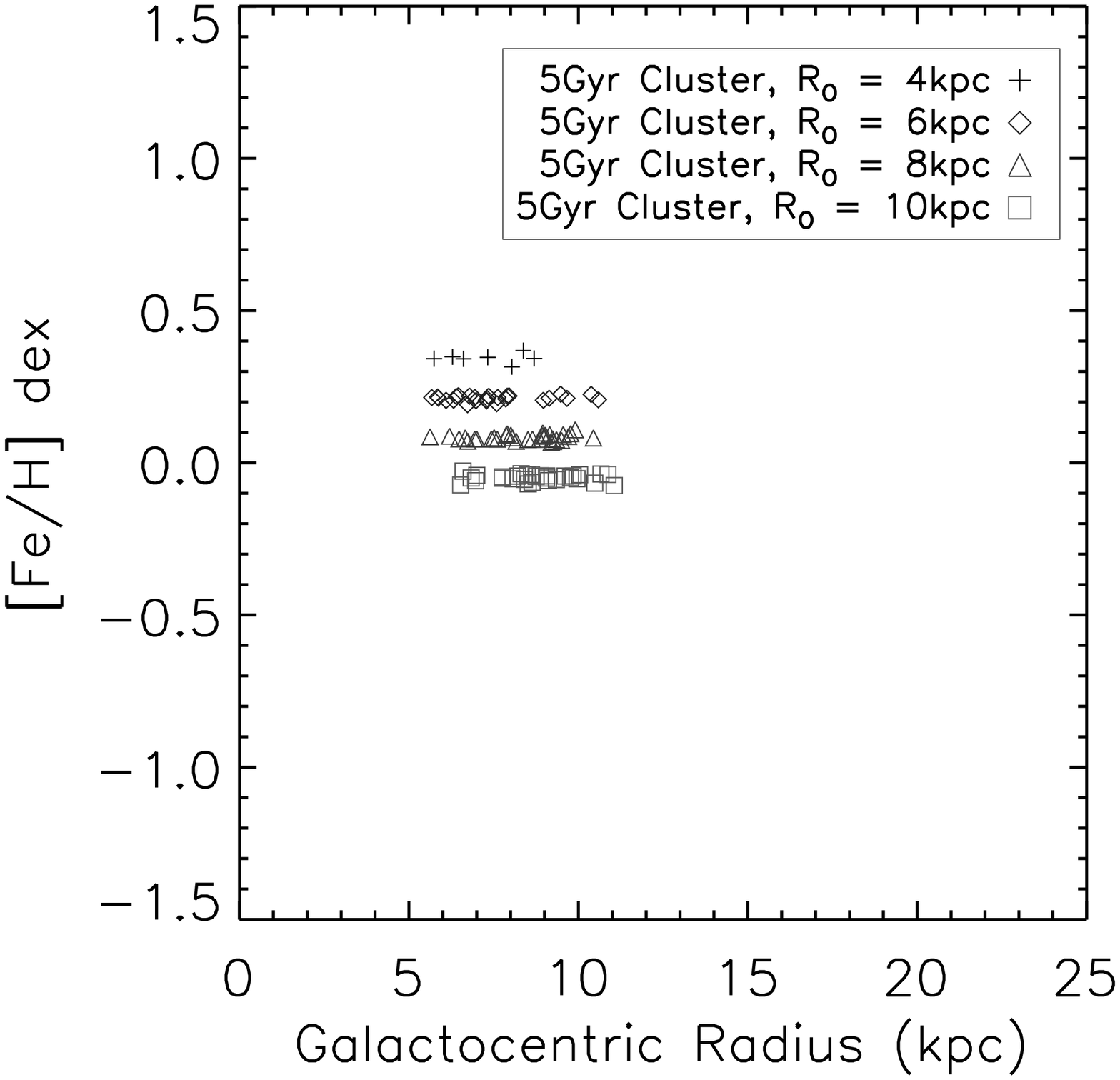}{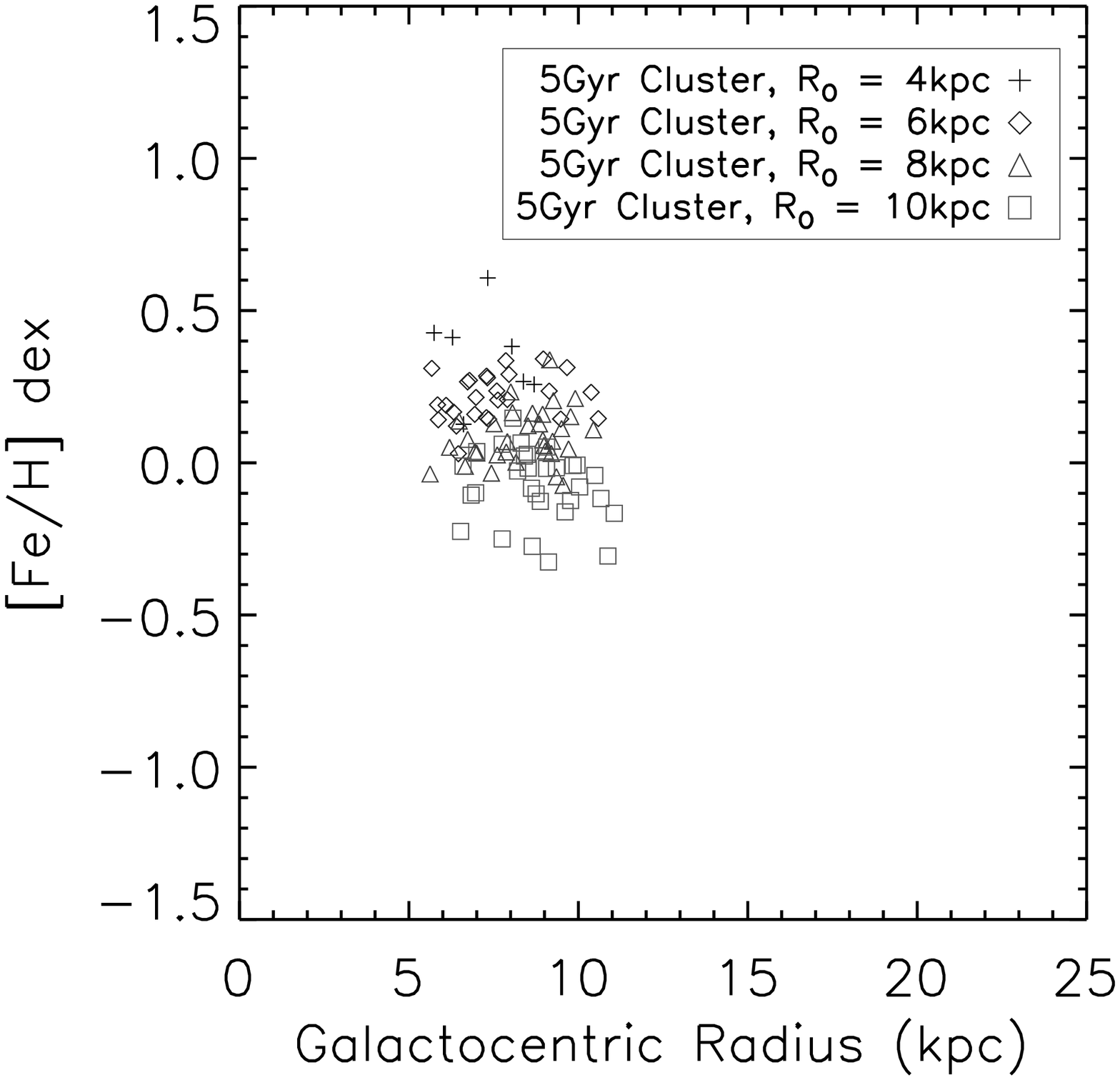}{Fig2}{Present-day
chemistry-radius distributions of four dissolved 250~M$_\odot$
stellar associations injected into the disk 5~Gyrs ago at 4 different
galactocentric radii, with chemical uncertainties of $\sigma_C=0.01$~dex
(left) and $\sigma_C=0.1$~dex (right).}

\begin{table}[!ht]
\caption{MST recovery accuracy and dissociation population percentages 
for different chemical abundance uncertainties.}
\smallskip
\begin{center}
{\small
\begin{tabular}{lrr}
\tableline
\noalign{\smallskip}
$\sigma_C$ (dex) & Recovery Accuracy (\%) & Dissociated Population (\%)\\
\noalign{\smallskip}
\tableline
\noalign{\smallskip}
\hspace{2mm}0.01 & 100\hspace{1.7cm} & 0\hspace{2cm} \\
\hspace{2mm}0.05 &  84\hspace{1.7cm} & 13\hspace{2cm} \\
\hspace{2mm}0.10 &  50\hspace{1.7cm} & 17\hspace{2cm} \\
\hspace{2mm}0.15 &  14\hspace{1.7cm} & 10\hspace{2cm} \\
\end{tabular}
}
\end{center}
\end{table}
\noindent

\vspace{-4mm}

\section{Conclusions}

\vspace{-4mm}
We employ a parametric form of the MST algorithm to search for now-dissolved
stellar associations, phase-mixed with a background exponential disk
of field stars.  Both the background sea of stars and the associations
themselves were chemically tagged with empirically-motivated
abundance patterns and a variety of chemical uncertainties and 
association injection radii were explored.  Not surprisingly, 
in the absence of chemical uncertainties, the association reconstruction
accuracy is high; equally unsurprising, the accuracy drops dramatically with
increasing abundance uncertainties.  While not meant to be exhaustive, 
the work presented here is a successful proof-of-concept.  Various 
weighting schemes, and non-parametric approaches, urgently need
to be explored, alongside more sophisticated multi-dimensional
group finding algorithms 
\citep{sj_12, m_13, m_14}.

\acknowledgements 
The support of our colleagues Jarrod Hurley, Guido Moyano Loyola, Simon 
Goodwin, and Rory Smith, is gratefully acknowledged.

\end{document}